\documentclass[letterpaper, 10pt]{article}

\usepackage{fullpage}
\usepackage{graphicx}
\usepackage[bf]{caption}
\usepackage{subfigure}
\usepackage{multicol}

\usepackage{amssymb}

 \usepackage{nopageno}

\newcommand{\beq}{\begin{equation}}
\newcommand{\eeq}{\end{equation}}
\newcommand{\bbar}{\begin{eqnarray}}
\newcommand{\eear}{\end{eqnarray}}

\newcommand{\I}{\mathrm{i}}

\begin{document}

\title{Cellular replication limits in the  Luria-Delbr\"uck mutation model}

\author{Ignacio A. Rodriguez-Brenes$\,^{1,2,*}$, Dominik Wodarz$\,^{2,1}$, and  Natalia L. Komarova$\,^{1,2}$}
\date{\small $^1$Department of Mathematics and $^2$Department of Ecology and Evolutionary Biology, University of California, Irvine CA 92697, U.S.A. $^*$Corresponding author. Email: iarodrig@uci.edu, Tel.: +1 949 824 2531.}


\maketitle

\begin{abstract}
Originally developed to elucidate the mechanisms of natural selection in bacteria, the Luria-Delbr\"uck model assumed that cells are intrinsically capable of dividing an unlimited number of times. This assumption however, is not true for human somatic cells which undergo replicative senescence. Replicative senescence is thought to act as a mechanism to protect against cancer and the escape from it is a rate-limiting step in cancer progression. Here we introduce a Luria-Delbr\"uck model that explicitly takes into account cellular replication limits in the wild type cell population and models the emergence of mutants that escape replicative senescence. We present results on the mean, variance, distribution, and asymptotic behavior of the mutant population in terms of three classical formulations of the problem. More broadly the paper introduces the concept of incorporating replicative limits as part of the Luria-Delbr\"uck mutational framework. Guidelines to extend the theory to include other types of mutations and possible applications to the modeling of telomere crisis and fluctuation analysis are also discussed.
\end{abstract}


\begin{center}
\line(1,0){460}
\end{center}

%
%

\section*{Introduction. }
~The Luria-Delbr\"uck experiment investigated whether mutations in bacteria arise spontaneously or as an adaptive response \cite{Luria:1943fk}. The answer led to a rich mathematical theory, with important applications in the calculation of mutation rates \cite{Kepler:2001fk}, the emergence of antibiotic-resistant microbes \cite{Johnson:2013zr}, the study of drug therapy-resistant cancer cells \cite{Haeno:2007kx,Diaz:2012nx} and cancer genetics \cite{Frank:2010dq,Komarova:2007oq}. Theoretical advances includude the analysis of the probability distributions \cite{Dewanji:2005ys, Angerer:2001ve}, asymptotic properties \cite{Kessler:2013tg}, numerical methods for fluctuation analysis \cite{Zheng:2002vn}, and the accuracy of estimates for the mutation rates \cite{Niccum:2012zr}. Extensions of the theory  include different cell cycle distributions and growth laws of wild type cells \cite{Angerer:2010qf,Tomasetti:2012hc}. For a review and more recent advances see \cite{Zheng:1999fk,Foo:2014bs}.

Currently an underlying assumption of the theory is that cells are capable of an unlimited number of divisions. This assumption is appropriate to model mutations in bacteria, it does not apply however, to the majority of cells in the human body. Normal human somatic cells are capable of a limited number of divisions, a phenomenon known as replicative senescence or Hayflick's limit \cite{HAYFLICK:1961ys}. Hence, when we consider the somatic evolution of human cells, it is fundamental to understand how replicative limits affect the emergence, dynamics, and distribution of mutant populations. Here, we present the first attempt to  address explicitly the role of replicative limits in the Luria-Delbr\"uck mutational framework.

Replicative senescence is linked to the shortening of telomeres during cell division, which are repetitive sequences of DNA found at the end of linear chromosomes \cite{Shay:2005vn}. Replication limits protect against cancer by limiting the size of a clonal cell population and  by reducing the possible number of cells divisions, when mutations typically occur. Cells can escape replicative senescence by expressing telomerase, an enzyme that extends telomere length \cite{Kim:1994fk}.

Essentially all human cancers acquire mechanisms to maintain telomere length, most often through high levels of telomerase expression (90\%) \cite{Kim:1994fk}, and less frequently through the alternative telomere lengthening pathway (ALT) (10\%) \cite{Bechter:2004fk}. The stage of tumor development at which cells start expressing telomerase is probably cancer-type-specific. Proliferation in telomerase negative cells after the inactivation of cell-cycle checkpoint pathways can lead to crisis, a phase characterized by widespread cell death and genome instability. Cells can emerge from crisis immortalized through telomerase activation \cite{Shay:2005vn}. This sequence of events might play an important role in breast cancer \cite{Chin:2004uq}. In other cancer types telomerase expression might occur at earlier stages \cite{Shay:2005vn}. If cancer originates in a telomerase positive stem cell, full progression towards malignancy might involve instead the up-regulation of telomerase activity. The finding in tumors of cells with stem cell characteristics has led to the concept of cancer stem cells (CSC). There is debate over the cell of origin of CSCs, whether they originate from normal stem cells or from differentiated cell types, which acquired stem cell characteristics. In multiple types of cancer there is evidence that the initiating mutations originate in cells with limited proliferative potential, such as progenitors (for a review see \cite{Visvader:2011uq}). 

Here, we consider mutations that allow cells to bypass replicative limits, and develop the model in terms of three formulations with different levels of randomness (the Luria-Delbr\"uck, Lea-Coulson and Bartlett formulations). We discuss several features of the mutant population including the expected value, variance, probability distribution, generating function and asymptotic properties. We also emphasize the comparison between the different statistics when replication limits are included with those from the classical model, where they are not.  We also investigate the probability of escaping replicative limits. Understanding this quantity is crucial to evaluate the effectiveness of replicative senescence as a tumor suppressor pathway. Finally, we outline two possible extensions of the theory and end by discussing applications of fluctuation analysis to estimate the rate of telomerase activation. 

%
%

\section*{Cellular replication limits}

To model replicative limits we assume that each cell has a  replication capacity $\rho\geq 0$. When a cell with replication capacity $\rho>0$ divides, it produces two daughter cells with replication capacities $\rho-1$. Cells with replication capacity $\rho=0$ become senescent and stop dividing (Fig.~1A). 

Let the rates of cell division and death be $a_{\rm div}$ and $a_{\rm die}$. If we denote the normalized division rate by $q= a_{\rm div}/(a_{\rm die}+a_{\rm div})$, then $a_{\rm div}= q(a_{\rm die}+a_{\rm div})$ and $a_{\rm die}= (1-q)(a_{\rm die}+a_{\rm div})$. Notice that in general $0\leq q \leq 1$ and if the cell population initially grows $q>0.5$. We can then express the model in terms of dimensionless units of time by making $a_{\rm die}+a_{\rm div}=1$. If $x_{\rho}(t)$ is the number of cells with replication capacity $\rho$ at time $t$ and $k$ is the maximum replication capacity, then the time evolution of the cell population is described by the system: 

\beq
\label{eq:ode_system}
\left\{ \begin{array}{lcccc}

 \dot{x}_k & = & -x_k\\
 
 \dot{x}_{k-1} & = &  2 q \,x_{k} & - & x_{k-1}\\
 
 \dot{x}_{k-2} & = &   2 q \,x_{k-1} & - &  x_{k-2}\\
 
&   \vdots &  & \\
 
  \dot{x}_0 & = & 2 q \,  x_1  & -  & (1-q)\, x_0\\

\end{array} \right. 
\eeq

We refer to propositions in the Supplementary Material with the letter {\it``S"} followed by a roman numeral.  If $X_{\rm tot}(t;q)= \sum_{\rho=0}^k x_\rho(t)$ and there is a single cell with replication capacity $k$ at $t=0$, from  {\it S1} we have: 

\beq
\label{eq:total}
X_{\rm tot}(t;q) = e^{-t} \left[ 2^k \sum_{n=k}^{\infty} \frac{(qt)^n}{n!} + \sum_{n=0}^{k-1} \frac{(2qt)^n}{n!}\right]
\eeq

Fig.~1B, plots the trajectory of $X_{\rm tot}(t;q)$ for different values of $k$. If $0.5<q<1$, the cell population first grows on account of the positive net growth rate ($a_{\rm div}>a_{\rm die}$). However, as time progresses more cells hit Hayflick's limit and stop dividing. When this occurs the cell population starts to decrease as most cells becomes senescent and eventually die. For an analysis of the stochastic version of [{\bf \ref{eq:ode_system}}] see \cite{Rodriguez-Brenes:2015ab}.

Since most mutations occur during cell division, from here on we focus our attention on the dividing fraction of the cell population  $X(t;q)= \sum_{j>0} x_j(t)$. In terms of the incomplete upper gamma function $\Gamma(k,t)= \int_t^\infty s^{k-1}e^{-s} {\rm d}s$, we find:

\beq
\label{eq:special_function}
X(t;q)= e^{(2q-1)t}\,\Gamma(k,2qt)/\Gamma(k)
\eeq

Eq.~{\bf \ref{eq:special_function}} can be modified to include different initial conditions. If $k$ is the maximum replication capacity found in the population at time $t = 0$, then the total number of dividing cells equals $e^{(2q-1)t} \sum_{j=1}^k   x_j(0)  \Gamma(j,2qt)/\Gamma(j)$. From here on we focus on populations arising from a single founding cell (Eq.~{\bf\ref{eq:special_function}}). Results for different initial conditions can be derived from the fact that the processes arising from each of the subpopulations $x_j(0)$ are independent of each other.

%
%

\section*{Luria-Delbr\"{u}k formulation} 

In the Luria-Delbr\"{u}k formulation \cite{Luria:1943fk} (hereafter referred to as LD formulation) the wild type cell population\footnote{Throughout the article we will write $X(t)$ to refer to a generic dividing wild type cell population (with or without replicative limits). When we refer to the specific function in [{\bf \ref{eq:special_function}}] we will write $X(t;q)$.}  is described as a deterministic function $X(t)$. The number of mutant cells at time $t$ originating from a single mutant born at time $s>0$ is also described by a deterministic function $Y(t,s)$. Mutations occur stochastically at a rate proportional to  $X(t)$, with a given per cell per unit of time mutation rate $\nu$.

In the classical formulation (without replication limits) $X(t)$ and  $Y(t,s)$ are  modeled as exponential functions of $t$. Here, we are interested in the number of telomerase positive mutants that arise from a growing population of cells restricted by Hayflick's limit. In this context $X(t)$ is a cell population with a limited replication capacity and is modeled with Eq.~ {\bf \ref{eq:special_function}}. Telomerase positive mutants are not subject to replicative limits and thus we keep  the description of $Y(t,s)$ as an exponential function. Throughout the paper we will compare the results of the model that assumes no replication limits, with the results of the model that includes them.

Mutations that occur during cell division affect the growth rate of the wild type cell population by removing from it one of the two daughter cells produced during mitosis. Indeed, if the rate of mutation per cell division is $\mu$, then for $0<j<k$ we now have $\dot x_j = (2q-q\mu)x_{j+1} - x_j$. If we make $\bar q= q(1-\mu/2)$, it follows from the structure of [{\bf \ref{eq:ode_system}}] that the number of dividing wild type cells is now $X(t;\bar q)= e^{(2\bar q-1)t}\Gamma(k,2\bar q)/\Gamma(k)$. Also, for this type of mutation we have $\nu= q\mu$.

We are interested in the total number of mutant cells  $Y(t)$. Following \cite{Crump:1974ab} we can write $Y(t)$  as filtered Poisson process [{\bf \ref{eq:filtered}}]. Here $M(t)$ equals the number of mutations in wild type cells by time $t$,  which is a Poisson process with rate $\nu X(t)$, and the $s_i$ are the times when the mutations occur.
\beq
\label{eq:filtered}
Y(t)= \sum_{i=1}^{M(t)} Y(t,s_i)
\eeq
Let $\psi(z;t,s) = {\rm E}[e^{\I z Y(t,s)}]$ be the characteristic function of $Y(t,s)$, then from the theory of filtered Poisson process \cite{Parzen:1999kx} we find that the characteristic function of $Y(t)$ equals:

\beq
\Psi(z;t)= \exp \left\{ \int_0^t \nu X(s) [\psi(z;t,s) -1] {\rm d}s \right\}
\eeq

$Y(t,s)= e^{\gamma(t-s) }$, where $\gamma$ is the growth rate of mutants. Hence, $Y(t,s)$ can be seen as a random variable with a degenerate distribution. We thus have $\psi(z;t,s)= e^{\I z e^{\gamma(t-s)}}$



%
%

\subsection*{Expected number of mutants in the LD formulation. }

%
%
%
%

From {\it S2} we have:

\beq
\label{eq:mean_generic}
{\rm E}[Y(t)] = \int_0^t \nu X(s) e^{\gamma(t-s)} {\rm d}s
\eeq

Using the parametrization in [{\bf\ref{eq:ode_system}}], when there are no replication limits $X(t)= e^{(2\bar q-1)t}$.  In this case from [{\bf\ref{eq:mean_generic}}] we recover the well known expression [{\bf\ref{eq:D_L-expected-nl}}]. Hereafter, we denote statistics for the models with no replications limits with the subscript ``nl" (no limits). For statistics of models with replication limits we use the subscript ``wl'' (with limits).

\beq
\label{eq:D_L-expected-nl}
{\rm E}[Y(t)]_{\rm nl}=  \left\{ \begin{array}{ll}
 \nu t e^{(2\bar q-1)t}  & \quad (\gamma = 2\bar q-1)  \\
  \frac{\nu}{(2\bar q-1)-\gamma} (e^{(2\bar q-1)t}-e^{\gamma t}) & \quad (\gamma \neq2\bar q-1)
       \end{array} \right. 
\eeq

When wild type cells are restricted by Hayflick's limit, we use Eq.~{\bf\ref{eq:special_function}} for $X(t)$. If mutants activate telomerase, the expected mutant cell population is given by Eq.~{\bf\ref{eq:avg_LC_wl_equal}} when $\gamma= 2\bar q-1$, and by Eq.~{\bf\ref{eq:avg_LC_wl_different}} when $\gamma \neq 2\bar q-1$ (Proposition {\it S3}).  Fig.~1C plots the trajectories of ${\rm E}[Y(t)]$ with and without limits for a case where $\gamma<2\bar q-1$.

\beq
\label{eq:avg_LC_wl_equal}
{\rm E}[Y(t)]_{\rm wl} =  \frac{\nu e^{(2\bar q-1)t} k}{2\bar q} + \nu( t-  k/(2\bar q))X(t;\bar q) -
\frac{\nu e^{-t}(2\bar qt)^k}{2\bar q\,\Gamma(k)}
\eeq

\beq
\label{eq:avg_LC_wl_different}
{\rm E}[Y(t)]_{\rm wl} = 
\frac{\nu}{(2\bar q-1)-\gamma} 
\left[
X(t;\bar q)
+ \left( \frac{2\bar q}{1+\gamma}\right)^k  \left( e^{\gamma t}- X(t; {\textstyle \frac{\gamma+1}{2}} ) \right)
 -  e^{\gamma t}
\right]
\eeq


%
%

\subsection*{Asymptotics  of the expected number of mutants.}

We examine the asymptotic behavior of ${\rm E}[Y(t)]_{\rm wl} $ when $t$ is large. We say two function $f$ and $g$ are asymptotically equivalent and write $f\sim g$ if $\lim_{t\rightarrow +\infty}f(t)/g(t) =1$. Given that $\lim_{t\rightarrow +\infty}X(t;q)= 0$, from [{\bf\ref{eq:avg_LC_wl_equal}}] and [{\bf\ref{eq:avg_LC_wl_different}}] as $t\rightarrow +\infty$:

\beq
\label{eq:avg_LC_asymp}
\begin{array}{ll}
{\rm E}[Y(t)]_{\rm wl} \sim \frac{\nu k}{2 \bar q} e^{(2\bar q-1)t}  & \quad (\gamma = 2\bar q-1)  \\
 {\rm E}[Y(t)]_{\rm wl} \sim\frac{\nu}{(2\bar q-1)-\gamma} \left( \left( \frac{2\bar q}{1+\gamma}\right)^k -1\right) e^{\gamma t} & \quad (\gamma \neq 2\bar q-1)
       \end{array} 
\eeq

Eq.~{\bf\ref{eq:LC_asymp_wl_nl}} gives the asymptotic relation between ${\rm E}[Y(t)]_{\rm nl}$ and ${\rm E}[Y(t)]_{\rm wl}$, where $A,B,C$ and $D$ are positive constants. From [{\bf\ref{eq:avg_LC_wl_equal}}] and [{\bf\ref{eq:avg_LC_wl_different}}], ${\rm E}[Y(t)]_{\rm wl} \leq {\rm E}[Y(t)]_{\rm nl} $ $\forall t\geq0$. However, depending on the relation between $\gamma$ and $2\bar q-1$, the asymptotic  behavior of ${\rm E}[Y(t)]_{\rm nl}$ and ${\rm E}[Y(t)]_{\rm wl}$, differs significantly. If $\gamma<2\bar q-1$, the growth of  ${\rm E}[Y(t)]_{\rm nl}$ is mostly driven by mutations in wild type cells, and thus ${\rm E}[Y(t)]_{\rm nl}/{\rm E}[Y(t)]_{\rm wl}$ increases as an exponential function of time. If $\gamma=2\bar q-1$, then this ratio increases as a linear function time. If $\gamma>2\bar q-1$, the growth of ${\rm E}[Y(t)]_{\rm nl}$ is mostly driven by self-replication of mutants and thus the ratio ${\rm E}[Y(t)]_{\rm nl}/{\rm E}[Y(t)]_{\rm wl}$ converges to $1/D$, where  $D= \left( 1-(2\bar q/(1+\gamma))^k \right) < 1$.

\beq
\label{eq:LC_asymp_wl_nl}
\begin{array}{ll}
{\rm E}[Y(t)]_{\rm wl} \sim A e^{-Bt} \,{\rm E}[Y(t)]_{\rm nl}  & \quad (\gamma < 2\bar q-1)  \\
 {\rm E}[Y(t)]_{\rm wl} \sim Ct^{-1} \, {\rm E}[Y(t)]_{\rm nl} & \quad (\gamma = 2\bar q-1) \\
{\rm E}[Y(t)]_{\rm wl} \sim D \, {\rm E}[Y(t)]_{\rm nl}& \quad (\gamma > 2\bar q-1)
       \end{array} 
\eeq

In the context of cancer it is also important to examine the expected total cell population, $ Z(t) = X_{\rm tot}(t,\bar q)+{\rm E}[Y(t)]$, as a measure of the overall tumor load. With limits $Z(t)_{\rm wl}$ grows without bound, but its size as a function of time is significantly reduced. For example, when the rates of cell division and death are equal for mutants and wild type cells $(\gamma= 2q-1)$, we have: $Z(t)_{\rm nl} = e^{(2q-1)t}$, and $Z(t)_{\rm wl} \sim ( 1-(1-\mu/2)^k)\,  e^{(2q-1)t}$. Fig.~1D illustrates these two cases. At  $t=0$ there are two cells, one with replicative limits and one without them. Initially, the populations  are almost identical, 
but eventually the wild type population with limits drops, as most of these cells become senescent. This trends continues until $Z(t)_{\rm wl}$ is composed almost entirely of mutants, at this point $Z(t)_{\rm wl}$ resumes growth as a roughly fixed fraction of $Z(t)_{\rm nl}$. Measurements of Hayflick's limit  typically yield a replication capacity for normal embryonic cells of 40-60 divisions \cite{HAYFLICK:1961ys}. If we use the midpoint value ($k=50$) and a mutation rate per cell division $\mu=10^{-9}$, then  $1-(1-\mu/2)^k \approx 2.5\times 10^{-8}$. This means that based on the expected number of cells, at later times $t$, the overall population which originated from a cell with replicative limits will be roughly eight orders of magnitude smaller than the population that originated from a cell without them. 

%
%

\subsection*{Variance in the LD formulation.} The formulas for the variance of the mutant population are: 1) With no limits Eq.~{\bf\ref{eq:LD-var-nl}}. 2) With limits Eq.~{\bf\ref{eq:LD-var-wl-diff}} for $2\gamma \neq 2\bar q -1$, and Eq.~{\bf\ref{eq:LD-var-wl-eq}} for  $2\gamma = 2\bar q -1$. (See {\it S4}). 

%
%

\beq
\label{eq:LD-var-nl} 
{\rm V}[Y(t)]_{\rm nl}=  \left\{ \begin{array}{ll}
\frac{\nu}{2\bar q-1-2\gamma}
\left(
e^{(2\bar q -1)t}
 -  e^{2\gamma t}
\right) & \textstyle \quad (2\gamma \neq 2\bar q-1)  \\
\nu t e^{2\gamma t} & \quad (2\gamma = 2\bar q-1)
       \end{array} \right. 
\eeq


\beq
\label{eq:LD-var-wl-diff}
\textstyle {\rm V}[Y(t)]_{\rm wl} = 
\frac{\nu}{2\bar q-1-2\gamma}
\left[ \scriptstyle
X(t;\bar q)
+ \left( \frac{2\bar q}{1+2\gamma}\right)^k  \left( e^{2\gamma t}- X(t; {\textstyle \frac{2\gamma+1}{2}} ) \right)
 -  e^{2\gamma t}
\right] 
\eeq

\beq
\label{eq:LD-var-wl-eq}
{\rm V}[Y(t)]_{\rm wl}=  \textstyle
\frac{\nu e^{(2\bar q-1)t} k}{2\bar q} + \nu( t-  k/(2\bar q))X(t;\bar q) -
\frac{\nu e^{-t}(2\bar qt)^k}{2\bar q\,\Gamma(k)} 
\eeq


Next, we analyze the asymptotics of ${\rm V}[Y(t)]$ as $t\rightarrow\infty$. Let  $R= {\rm V}[Y(t)]_{\rm nl}/{\rm V}[Y(t)]_{\rm wl}$. For $t$ large we find: If $2\gamma< 2\bar q-1$, $R$ increases as an exponential function of $t$; if  $2\gamma= 2\bar q-1$, $R$ increases as linear function of time;  if $2\gamma>2\bar q-1$, $R$ converges to a fixed positive number. Interestingly this can lead to situations for which $ {\rm V}[Y(t)]_{\rm nl}/{\rm V}[Y(t)]_{\rm wl}$ grows at a substantially different rate than $ {\rm E}[Y(t)]_{\rm nl}/{\rm E}[Y(t)]_{\rm wl}$. One important case occurs when $\gamma= 2q-1$. In this case $\gamma>2\bar q -1$ and thus  ${\rm E}[Y(t)]_{\rm nl}/{\rm E}[Y(t)]_{\rm wl}$ and $R$ both converge to positive numbers. However, if $\mu$ is very small, $\gamma$ can be very close to $2\bar q-1$ (a value where ${\rm E}[Y(t)]_{\rm nl}/{\rm E}[Y(t)]_{\rm wl}$  grows as a linear function of time). In this case the convergence of $ {\rm E}[Y(t)]_{\rm nl}/{\rm E}[Y(t)]_{\rm wl}$ will be very slow. As a consequence, for the number of cells typically considered during experiments, the growth of the ratio of the expected values is essentially linear (Fig.~1E). In contrast, when $\gamma= 2q-1$, ${\rm V}[Y(t)]_{\rm nl}/{\rm V}[Y(t)]_{\rm wl}$ converges rapidly to the limit $ ( 1- (2\bar q/(4q-1))^k )^{-1}$ (Fig.~1F). Interestingly this limit is close to one for a wide range of biologically relevant parameters. For example, for $k=50$, $\mu=10^{-9}$ and $q=0.55$, which is only 10\% larger than the zero net growth condition $q=0.5$, this limit is approximately 1.01 and it decreases with larger $q$. Hence, for a wide range of biologically relevant parameters,  when $\gamma=2q-1$, the variance in the number of mutants should be roughly the same with or without limits; while (for a biological meaningful range of cells) the ratio $ {\rm E}[Y(t)]_{\rm nl}/{\rm E}[Y(t)]_{\rm wl}$ increases steadily with time (Figs.~1E and 1F).

%
%

\subsection*{Probability that no mutants are present.}

Once the first mutation occurs in the LD formulation, the existence of mutants is guaranteed at all later times. Hence, to calculate $P_0(t)$, the probability that no mutants are present at time $t$, it is sufficient to consider the random variable $B(t)$, defined as the cumulative number of mutations in the interval $[0,t]$. If $H$ is the right continuous Heaviside function, then we can write $B(t)$ as a filtered Poisson process: $B(t)= \sum_{i=1}^{M(t)}{H(t-s_i)}$ (see [{\bf\ref{eq:filtered}}] for notation). Since  the probability generating function (p.g.f.) of $H(t-s)$ is $z^{H(t-s)}$, the p.g.f.~of $B(t)$ equals: 

\beq
\label{eq:pgf_B}
	\Phi_{B}(z;t)= \exp \left\{ \int_0^t \nu X(s) [z^{H(t-s)}-1]{\rm d}s \right\}
\eeq

The probability that no mutations take place in the interval $[0,t]$ is $\Phi_{B}(0;t)  = P_0(t)$. We thus find:


\beq
\label{eq:P0_nl}
	P_0(t)_{\rm nl} = \exp \left\{ \frac{\nu}{2\bar q-1} \left( 1-e^{(2\bar q-1)t} \right) \right\}
\eeq



\beq
\label{eq:P0_wl}
	P_0(t)_{\rm wl}= \exp \left\{ \frac{\nu}{2\bar q-1}
	\left(
	\frac{(2\bar q)^k \Gamma(k,t)}{\Gamma(k)} 
	-X(t;\bar q) -(2\bar q)^k + 1
	\right)
	\right\}
\eeq

The probability that no mutations ever take place is $\lim_{t\rightarrow \infty} \Phi_B(t;0)$, which we write as $P_0(\infty)$. 
The first two terms inside the round brackets of [{\bf\ref{eq:P0_wl}}] vanish as $t\rightarrow\infty$. Hence, $P_0(\infty)_{\rm wl}$ is given by [{\bf\ref{eq:P0inf_wl}}]. This quantity is especially significant, as $1-P_0(\infty)$ can be interpreted as the probability that an abnormally growing colony of cells eventually escapes the cancer protecting mechanism of replicative senescence.

\beq
\label{eq:P0inf_wl}
P_0(\infty)_{\rm wl}= \exp \left\{ -\nu \frac{(2\bar q)^k-1 }{2\bar q -1} \right\}
\eeq

With no limits the probability that there will never be a mutation is zero (i.e., $P_0(\infty)_{\rm nl} = 0$), which follows from [{\bf\ref{eq:P0_nl}}]. Fig.~2A plots $P_0(t)$ as a function of time with and without limits for various values of $k$. The limiting values $P_0(\infty)_{\rm wl}$ can be inferred from the horizontal asymptotes of the plots. Other ways of calculating $P_0(\infty)$ are discussed below (see Fig.~2C).

%
%

\section*{Lea-Coulson formulation}

The Lea-Coulson formulation (hereafter referred to as the LC formulation) differs from the LD formulation in the modeling of $Y(t,s)$
as a linear birth-death process with constant parameters. Thus, $\phi(z;t,s)$ the p.g.f. of $Y(t,s)$, is given by [{\bf\ref{eq:pgs_LC_bd}}], where $\alpha$ and $\beta$ are respectively the birth and death rates of mutants.  $Y(t,s)$ is then a filtered Poisson process with p.g.f. $\Phi(z;t)= \exp \left\{ \int_0^t \nu X(s) [\phi(z;t,s) -1] {\rm d}s \right\}$.

%
%
%
%
%
%
%
%

\beq
\label{eq:pgs_LC_bd}
	\phi(z;t,s)= \frac{\beta(1-z)-(\beta-\alpha z)e^{-(\alpha-\beta)(t-s)} }{\alpha(1-z)-(\beta -\alpha z)e^{-(\alpha-\beta)(t-s)}  }
\eeq

%
%

\subsection*{Statistics in the LC formulation.}

${\rm E}[Y(t)]$ is same the LD and LC formulations. If we indicate the formulations with a superscript, then with and without the limits the variances satisfy the identity [{\bf\ref{eq:var-relation}}] (Propositions {\it S5}).

\beq
\label{eq:var-relation} \textstyle
{\rm V}[Y(t)]^{\scriptscriptstyle \rm (LC)}=
{\rm E}[Y(t)] +
\frac{2\alpha}{\alpha-\beta} \left( {\rm V}[Y(t)]^{\scriptscriptstyle \rm (LD)} - {\rm E}[Y(t)] \right)
\eeq

From [{\bf\ref{eq:var-relation}}], the observation that for $\beta>0$, ${\rm V}[Y(t)]^{\scriptscriptstyle \rm (LC)}_{\rm nl}$ increases with $\alpha$, even when the net growth rate ($\alpha-\beta$) is fixed \cite{Dewanji:2005ys}, also  holds for ${\rm V}[Y(t)]^{\scriptscriptstyle \rm (LC)}_{\rm wl}$. When $\beta=0$, it has also been observed that ${\rm V}[Y(t)]^{\scriptscriptstyle \rm (LC)}_{\rm nl}$ is roughly twice the size of ${\rm V}[Y(t)]^{\scriptscriptstyle \rm (LD)}_{\rm nl}$ when the growth rates of wild type cells and mutants are equal \cite{Zheng:1999fk}. From [{\bf\ref{eq:var-relation}}], this observation holds in the model with limits for $t$ large, even when the growth rates are different. This comes from the fact that ${\rm E}[Y(t)]_{\rm wl} \sim A e^{(\alpha-\beta)t}$ and $ {\rm V}[Y(t)]_{\rm wl}^{\scriptscriptstyle \rm (LD)} \sim B e^{2(\alpha-\beta) t}$ for some $A,B\in \mathbb{R}^+$. Moreover, with and without limits, $ {\rm V}[Y(t)]^{\scriptscriptstyle \rm (LD)}  \sim C {\rm V}[Y(t)]^{\scriptscriptstyle \rm (LC)} $ for some $C\in\mathbb{R}^+$. Hence, the basic asymptotic relations derived for the expected values and variances in the LD formulation hold for the LC formulation. Explicit solutions for the LC variances can be written down for any choice of parameters $q,\mu,\alpha \mbox{ and } \beta$ by substituting in [{\bf\ref{eq:var-relation}}] the appropriate Eqs.~{\bf\ref{eq:D_L-expected-nl}}-{\bf\ref{eq:avg_LC_wl_different}} and {\bf\ref{eq:LD-var-nl}}-{\bf\ref{eq:LD-var-wl-eq}}.

 %
 %

 \subsection*{Distribution.}
 
The series representation $ \log \Phi(z;t) = \sum q_n(t) z^n$ can be used to calculate the probability density of $Y(t)$ numerically \cite{Ma1992:ab}, where the $q_n(t)$ are given by [{\bf\ref{eq:dist-2}}] (See {\it S6} for the validity of this expression in the model with limits).

  
\beq
  \label{eq:dist-2}
q_n(t) =
\int_0^t \nu X(s) \left(
\frac{\alpha-\beta}{\alpha-\beta e^{-(\alpha-\beta)(t-s)} }
\right)^2
\left( \frac{\alpha-\alpha e^{-(\alpha-\beta)(t-s)} }{\alpha-\beta e^{-(\alpha-\beta)(t-s)}  }  \right)^{n-1}
e^{-(\alpha-\beta)(t-s)}
\,{\rm d}s
\eeq
 
\noindent The probabilities $p_n(t)= {\rm P}(Y(t)=n)$ can be calculated recursively, from $p_0(t)= e^{q_0(t)}$ and [{\bf\ref{eq:recursive}}] (see lemma 2 in \cite{Zheng:1999fk}).
 
 \beq
 \label{eq:recursive}
 	p_n(t)= n^{-1} \sum_{j=0}^{n-1} (n-j) q_{n-j}(t) p_j(t) \quad (n\geq 1)
 \eeq

In Fig.~2B we integrate numerically the $q_n$ and use [{\bf\ref{eq:recursive}}] to compute the density of $Y(t)$, for a fixed value of $t$. One curve plots the distribution of a mutant population that arises from a single wild type cell with no replicative limits. The other three curves trace the distributions that arises from founding cells with different replication capacities. Note that with lower replication capacities $k$, both the mode and the spread of the distributions decrease. Conversely, as $k$ increases the distributions progressively resemble that of the model without limits. Indeed, when we make $k=50$ (not shown) the plots with and without limits are indistinguishable. It is important to note however, that with time any distribution corresponding to a fixed value of $k$ will progressively diverge from the distribution without limits.

%
%

\subsection*{Probability that no mutants are present.} Since $\Phi(0;t) = P_0(t)$ with  replicative limits we have:


\beq
\label{eq:p0-lc-wl-new}
P_0(t)_{\rm wl} = \exp \left\{ -\frac{\nu (\alpha-\beta)}{ \alpha \,\Gamma(k)} \int_0^t \frac{e^{(2\bar q-1)s} \,\Gamma(k, 2\bar q s)}{1 - \frac{\beta}{\alpha} e^{-(\alpha-\beta) (t-s)}}  \, {\rm d}s \right\}
\eeq


Call $A= -\beta/ \alpha$ and $y= e^{-(\alpha-\beta) (t-s)} $, then $\left[ 1 - \frac{\beta}{\alpha} e^{-(\alpha-\beta) (t-s)} \right]^{-1} = 1/(1+A y)$. We want to expand $1/(1+Ay)$ as a Taylor series around $y=0$. Note that the only singularity of this function occurs at $y=-1/A$; hence, it is easy to prove that the expansion $1/(1+Ay)= \sum_{n=0}^{+\infty} (-A)^n y^n$ is valid for $|y|<-1/A$. Now, given our definition of $y$, if $0\leq s \leq t \: \Rightarrow \: 0 < y \leq 1$. Furthermore, $1<-1/A  \Leftrightarrow  1/2<q$. Hence, if the net growth of mutant population is positive ($1/2<q$), then $0<y< -1/A$ and thus the Taylor expansion for $1/(1+Ay)$ is valid for $0\leq s \leq t$. We have then:

\beq
\textstyle
 \left[ 1 - \frac{\beta}{\alpha} e^{-(\alpha-\beta) (t-s)} \right]^{-1} = \sum_{n=0}^{\infty} \left( \frac{\beta}{\alpha} \right)^n
 e^{-(\alpha-\beta)n(t-s)} 
\eeq

If we denote the integral  in [{\bf\ref{eq:p0-lc-wl-new}}] as $I(t)$, we have:

\beq
\label{eq:I(t)-1}
\textstyle
I(t) = \int_0^t e^{(2\bar q-1)s} \Gamma(k, 2 \bar q s) \sum_{n=0}^{\infty} \left( \frac{\beta}{\alpha} \right)^n
 e^{-(\alpha-\beta)n(t-s)} \, {\rm d}s
\eeq

Using Weierstrass M-test it is straightforward to prove that the series in [{\bf\ref{eq:I(t)-1}}] converges uniformly in $[0,t]$. Hence, we can interchange the order of integration and summation in the previous expression. If we do this, we get the expression [{\bf\ref{eq:expression_I(t)}}] for $I(t)$, where $h_n(s)= e^{((\alpha-\beta)n +2\bar q-1 )s} \Gamma(k, 2\bar q s)$:

\beq
\label{eq:expression_I(t)}
\textstyle
I(t) = \sum_{n=0}^{\infty} e^{-(\alpha-\beta)nt}  \left( \frac{\beta}{\alpha} \right)^n \int_0^t h_n(s) {\rm d}s
\eeq 

We can calculate $\int_0^t h_n(s) {\rm d}s$ using Eq.~A7 in the appendix and write\footnote{We can use [A7] when $1-(\alpha-\beta)n\neq 0$; in case of equality, $e^{-(\alpha-\beta)nt}\int_0^t h_n(s){\rm d}s = 1/(2\bar q) [e^{-t}(2\bar q t)^k/k + X(t;\bar q)\Gamma(k)-e^{-t}\Gamma(k)]$, which is bounded for $t\geq 0$ and goes to zero as $t$ goes to infinity (note that this situation can occur for at most one $n>0$).}:

\noindent $I(t)= \sum_n \left(\beta/\alpha\right)^n \left(a_n(t)+b_n(t)+c_n(t)\right)/d_n$. If we denote $\delta_n = 1-(\alpha-\beta)n$, we have $a_n(t)= X(t;\bar q ) \Gamma(k)$, $ b_n(t)= - ( 2\bar q/\delta_n)^k X(t;\delta_n/2)\Gamma(k)$, $ c_n(t)= e^{(\delta_n-1)t} [( 2\bar q/\delta_n)^k -1)] \Gamma(k) $ and $d_n= (\alpha-\beta)n +2\bar q-1$.

We are interested in $\lim_{t\rightarrow +\infty}I(t)$. To calculate this limit we will first prove that we can interchange the order of the limit and the summation. First, given that $\lim_{t\rightarrow +\infty} X(t;\bar q)= 0$ and $\delta_n-1 \leq 0 $, it is clear that $a_n(t)$ and $c_n(t)$ are uniformly bounded for $t\geq0$. Similarly, we can prove that $b_n(t)$ is uniformly bounded by $(2\bar q /\min |\delta_n|)^k\Gamma(k)$. Now, let $f_t(n)= (\beta/\alpha)^n \left(a_n(t)+b_n(t)+c_n(t)\right)/d_n$, then $I(t)$ is the integral of $f_t(n)$ with respect to the counting measure on $\mathbb{N}$. Since $a_n(t),b_n(t)$ and $c_n(t)$ are uniformly bounded, there is a constant  $M$, such that $|f_t(n)|\leq (\beta/\alpha)^n M/d_n$ $\forall n \in \mathbb{N}$.  Hence, given that $\sum_n (\beta/\alpha)^n M/d_n$ converges (i.e., the integral with respect to the counting measure exists), we can apply Lebesgue's dominated convergence theorem and interchange the order of the limit with respect to $t$ and the summation over $n$. We have then: $\lim_{t\rightarrow +\infty} I(t) = \sum_n \lim_{t\rightarrow+\infty} f_t(n)$.

From the definitions of $a_n(t), b_n(t) \mbox{ and } c_n(t)$ it is clear that when $t$ goes to infinity, $a_n(t)$ and $b_n(t)$ go to zero for $n\geq0$, and  $c_n(t)$ goes to zero for $n\geq1$. Hence: $\lim_{t\rightarrow +\infty}I(t) =  \frac{\left(2\bar q\right)^k-1}{2\bar q -1}\: \Gamma(k)$.


\noindent Substituting the last expression for $I(t)$ into [{\bf\ref{eq:p0-lc-wl-new}}], we find:

\beq
\label{eq:P0inf_LC_wl_final}
	P_0(\infty)_{\rm wl} = \exp\left\{ 
		-\nu \: \frac{\alpha-\beta}{\alpha} \:\frac{\left(2\bar q\right)^k-1}{2\bar q -1}
	\right\}
\eeq

If we write erl as an abbreviation for ``escaping replication limits'', the relationship between the probability of escaping replication limits in the LD and LC formulations is given by Eq.~{\bf\ref{eq:Perl_LC_vs_LD}} (see Fig.~2C). When there is no mutant death $(\beta=0)$ we have $P({\rm erl})^{\scriptscriptstyle \rm (LC)}= P({\rm erl})^{\scriptscriptstyle \rm (LD)}$. When $\beta$ is close to $\alpha$, $P({\rm erl})^{\scriptscriptstyle \rm (LD)}$ is much larger than $P({\rm erl})^{\scriptscriptstyle \rm (LC)}$. 
 
\beq
\label{eq:Perl_LC_vs_LD}
	P({\rm erl})^{\scriptscriptstyle \rm (LC)} = 1-\left( 1-P({\rm erl)}^{\scriptscriptstyle \rm (LD)} \right)^{1-\frac{\beta}{\alpha}} 
\eeq

%
%

\section*{Bartlett formulation} 

This is the fully stochastic formulation of the Luria-DelbrŸck model, where both $X(t)$ and $Y(t,s)$ are birth-death processes \cite{Bartlett:1978ly}. When there are no  limits, differential equations for the joint p.g.f., $\Phi_{X,Y} (t;z,w)$, exist \cite{Dewanji:2005ys}. Eq.~{\bf\ref{eq:joint_pgf}} gives the solution for $\Phi$  when there is no death ($q=1,\beta=0$), the birth rates are equal ($q=\alpha$) and there is a single wild type cell at $t=0$ (i.e., $\Phi(0,z_1,z_2)=z_1$) \cite{Bartlett:1978ly}. Recently a closed solution for $\Phi$ in the general case has been found in terms of a complex set of formulae \cite{Antal:2011ab}.

 \beq
\label{eq:joint_pgf}
\Phi(t,z,w)= -\frac{z w}{
 z (e^t (-1 + w) - w) + e^t (z - w) (1 - w + e^{-t} w)^\mu}
 \eeq

To include limits in Bartlett's model we consider the subpopulations $x_{\rho}(t)$, equal to the number of cells with replication capacity $\rho$ at time $t$. This approach defines a continuous-time Markov process. If  $P_{i_0,\ldots,i_k,l}(t) = P(x_0(t)=i_0,\dots,x_k(t)=i_k,Y=l)$, then we can write the Kolmogorov forward equations for the states $P_{i_0,\ldots,i_k,l}(t)$ and from them deduce a differential equation for the joint p.g.f.~of $(x_0,\ldots,x_k,Y)$ [{\bf\ref{eq:diff_eq_bartlett_wl}}]. (Eq.~{\bf\ref{eq:diff_eq_bartlett_wl}} assume a zero death rate. For the general equation and details on the derivation see {\it S7}) A detailed analysis of [{\bf\ref{eq:diff_eq_bartlett_wl}}] is the subject of  planned future research.




 \beq
 \label{eq:diff_eq_bartlett_wl}
 \frac{\partial \Phi}{\partial t}=  
 \sum_{j=1}^k \left[  (1-\mu) z_{j-1}^2 + \mu z_{j-1} w - z_j  \right] \frac{\partial \Phi}{\partial z_j}
 +
 \left( \alpha w^2 - \alpha w\right) \frac{\partial \Phi}{\partial w}
 \eeq

If there is no death, with  limits, the total number of divisions that eventually occurs  is $2^k -1$. Hence,  the probability that a mutant colony is never established is the same as the probability that there will never be a mutation. We thus have: $P_0(\infty)_{\rm wl} = (1-\mu)^{2^k-1}$. When $q<1$, the PDE for $\Phi$ can be solved numerically for small values of $k$; however, when $k$ is large this is computationally unfeasible. Instead, we use a hybrid stochastic-deterministic algorithm (developed in \cite{Rodriguez-Brenes:2015ab}) to simulate the dynamics of wild type cells and calculate the distribution of the total number of divisions. If $M$ is the total number of mutations and $r=(1-q)/q$, then using conditional probabilities and the distribution of the total number of divisions to calculate $P(M>0)$, we find:.

\beq
\label{eq:P(erl)_bounds}
	(1-r) \,P(M>0) \leq P({\rm erl)}^{\scriptscriptstyle \rm (B) } \leq P(M>0)
\eeq

Fig.~2C plots $P({\rm erl)}^{\scriptscriptstyle \rm (LD)}$, $P({\rm erl)}^{\scriptscriptstyle \rm (LC)}$ and $P(M>0)$. Note that a significant deviation can occur between $P(M>0)$ and $P({\rm erl})^{\scriptscriptstyle \rm (LC,LD)}$ when the latter probabilities are close to one (e.g., in Fig.~2C when $k=42$ and $q=0.35$). In these cases the maximum replication capacity is large, resulting in large values for $P({\rm erl})^{\scriptscriptstyle \rm (LC,LD)}$. This occurs in part, because the LC and LD formulations do not account for the possibility of early stochastic extinction of the wild type cell populations.

%
%

\section*{Discusion}

The Luria-Delbr\"uck theory as a framework to study mutation processes, has been a very effective tool with a wide variety of applications. From its inception (with its origins in bacterial modeling) an implicit assumption of the theory has been that cells are capable of an unlimited number of divisions. This assumption however, is not true for normal human somatic cells which are subject to Hayflick's limit. In this article we presented a Luria-Delbr\"uck model, which takes into account replicative limits in the wild type cell population and the emergence of mutants that escape replicative senescence

In a broader context, the ideas presented in this paper can form the basis to develop a more general Luria-Delbr\"uck framework for cells with a limited replication capacity.  Possible extension of the theory include the study of mutations that do not allow for unlimited cell replication. An LD type formulation could be developed for this purpose using a deterministic function for the growth of the mutant clones; an LC type formulation could follow from an adaptation of the methods in \cite{Luebeck:1991fk}. This type of mutation is related to a sequence of events that can play an important role in certain types of cancer. Proliferating cells with intact cell-cycle checkpoint pathways typically cease to divide once telomeres reach a critically short size (a stage called replicative senescence or M1 stage). Cells with defective cell-cycle checkpoint pathways might bypass M1 while their telomeres continue to shorten eventually leading to crisis (or M2 stage). Crisis is characterized by uncapped telomeres causing chromosome breakage-fusion-bridge cycles and widespread apoptosis \cite{Shay:2005vn,Chin:2004uq}. The rare cell which escapes crisis might then harbor genomic aberrations acquired during the period of genomic instability \cite{Chin:2004uq}. This sequence of events is akin to a two-step mutation process in which replicative limits play a central role. A Luria-Delbr\"uck type of approach to model a two-step mutation process without replication limits can be found in \cite{Luebeck:1991fk,Dewanji:1989ve}.

Another extension of the model is the development of fluctuation tests to measure the rate of escape from replicative senescence. The adaption of the so-called $P_0$ method is straightforward and comes directly from the inversion of the formulas for $P_0(\infty)$. Other methods based on the numerical computation of the probability distribution of mutants  can be applied to estimate the rate of immortalization \cite{Zheng:2002vn}. However, given the breadth of the subject a thorough analysis of the topic falls outside the scope of this paper. One important thing to mention is the scope of applicability of fluctuation tests in an experimental setting. Although the spontaneous immortalization of human cells in culture has been documented \cite{Diebold:2003zr}, it is a rare event not amenable to statistical analyses. However, human cells transfected with tumor viruses, such as SV40, or from patients with genetic disorders that carry increased cancer risk, such as Li-Fraumeni syndrome, immortalize in culture at detectable and reproducible frequencies \cite{Shay:1993vn,Herbert:2001uq}. In these cells the calculation of immortalization rates has clinical applications, as they can be used to assess the effectiveness of chemopreventive and antitelomerase agents \cite{Herbert:2001uq}.

%
%

{\footnotesize

\bibliographystyle{ieeetr}

}

%
%

\newpage

\section*{Figure Legends}

\bigskip

{
\noindent{\bf Figure 1.} {
A) Each cell has a  replication capacity $\rho\geq 0$. When a cell with replication capacity $\rho>0$ divides, it produces two daughter cells with replication capacities $\rho-1$. Cells with replication capacity $\rho=0$ become senescent and stop dividing.
 B) Total number of wild type cells $X_{\rm tot}$ vs time. In each plot there is a single cell with replication capacity $k$ at time $t=0$, normalized division rate $q=0.7$ (Eq.~{\bf\ref{eq:total}}).
  C) Expected number of mutant ${\rm E}[Y(t)]$, with limits (solid lines) and without limits (dashed lines). Parameters: $q=0.6$, $k=20$, mutation rate $\mu=10^{-9}$ and mutant growth rate $\gamma= 0.15$ (Eqs.~{\bf\ref{eq:D_L-expected-nl}} and {\bf\ref{eq:avg_LC_wl_different}}).
  D) Expected total cell population $Z(t)=X_{\rm tot}(t)+{\rm E}[Y(t)]$ as a function of time. Parameters for panels D, E and F: $q=0.55,\,k=50,\,\mu=10^{-9}, \mbox{ and } \gamma= 2 q-1$.
   E) Ratio of the expected numbers of mutants with and without limits. Here, $\gamma= 2q-1$, and thus ${\rm E}[Y(t)]_{\rm nl}/{\rm E}[Y(t)]_{\rm wl}$ converges to a fraction less than one (see text). However, because $\gamma$ is very close to $2\bar q-1$, where  ${\rm E}[Y(t)]_{\rm wl} \sim Ct^{-1} \, {\rm E}[Y(t)]_{\rm nl}$, the convergence is slow and thus, for the range of cells typically observed during experiments, the growth of ${\rm E}[Y(t)]_{\rm nl}/{\rm E}[Y(t)]_{\rm wl}$ is essentially linear.
    F) Ratio of the variance of the mutants populations with and without limits.}\\

\bigskip

\noindent{\bf Figure 2.} {A) Probability that there are no mutants at time $t$ (LD formulation). The horizontal asymptotes of the plots indicate the limiting values $P_0(\infty)_{\rm wl}$. Parameters: $q=1$ and $\mu= 10^{-9}$. B)  Probability density of the number of mutants (LC formulation). Parameters taken from Figs.~1 of \cite{Dewanji:2005ys,Zheng:1999fk}:  $t=6.7$, $\mu=10^{-7}$, $\alpha=10$, $\beta=7.5$ and $q=2$. C) Probability of escaping replication limits $P({\rm erl})$ in the LC and LD formulations, and $P(M>0)$, the probability that there is at least one mutation during the processes in Bartlett's formulation. Parameters: $\alpha=q$, $\beta=1-q$ and $\mu=10^{-9}$.}
}

\end{document}


\maketitle

\begin{center}
\line(1,0){460}
\end{center}

In this appendix reference to equations that appear in the main text are labeled with arabic numerals, e.g. (1). Equations that appear in this appendix are labelled with the letter ``A'' followed by an arabic numeral, e.g. (A1).\\

\subsection*{Cellular replication limits}

%
%

\bigskip

\noindent {\it {\sc Proposition S1}. ``If the $x_j(t)$ are defined by system {\rm [{\bf 1}]}, $x_k(0)=1$, and $x_j(0)=0$ for $j<k$, then $ \sum_{\rho=0}^k x_\rho(t) = e^{-t} \left[ 2^k \sum_{n=k}^{\infty} \frac{(qt)^n}{n!} + \sum_{n=0}^{k-1} \frac{(2qt)^n}{n!}\right]$.}''\\


\noindent{\it Proof:}  From {\rm [{\bf 1}]} we have  $\dot{x}_{j}  =   2 q \,x_{j+1}  -  x_{j}$. Hence, using the initial conditions and integration by parts we find:  $x_j(t)= e^{-t}\int_0^t 2q \,x_{j+1}(s)\,e^s {\rm d}s$ for $0<j<k$, and $x_0(t)=e^{(q-1)t} \int_0^{t} 2q\, x_1(s) e^{(1-q)s} {\rm d}s$.  Given that $x_k(t)= e^{-t}$ we can prove by induction that $x_j(t)= e^{-t} \frac{(2qt)^{k-j}}{(k-j)!}$ for $j>0$ and $x_0(t)= 2^ke^{-t} \sum_{n=k}^\infty \frac{(qt)^n}{n!}$. The total cell population $ X_{\rm tot}(t;q) = \sum_{\rho=0}^k x_\rho(t)$ is then given by:

\beq
\label{eq:total}
X_{\rm tot}(t;q) = e^{-t} \left[ 2^k \sum_{n=k}^{\infty} \frac{(qt)^n}{n!} + \sum_{n=0}^{k-1} \frac{(2qt)^n}{n!}\right]
\eeq
{ \flushright $\qedsymbol$\\ }

\subsection*{Luria-Delbr\"uck formulation}

%
%

\bigskip

\noindent {\it {\sc Proposition S2.} ``In the Luria-Delbr\"uck formulation the expected number of mutants
 ${\rm E}[Y(t)]= \int_0^t \nu X(s) e^{\gamma(t-s)} {\rm d}s$.}''\\
 

\noindent {\it Proof:} From [{\bf 5}] the characteristic function of the number of mutants $Y(t)$ is $\Psi(z;t)= \exp \left\{ \int_0^t \nu X(s) [\psi(z;t,s) -1] {\rm d}s \right\}$, where $\psi(z;t,s)= e^{\I z e^{\gamma(t-s)}}$. Let $f(s,z)$ and $g(s)$ be two functions with continuous second order partial derivates, then:

\beq
\label{eq:3}
\frac{\rm d}{{\rm d}z} \exp\left[\int_0^t g(s)f(s,z) {\rm d}s \right] = e^{\int_0^t g(s)f(s,z){\rm d}s} \int_0^t g(s) \frac{\partial f(s,z)}{\partial z} {\rm d}s 
\eeq

If we make $f(s,z)= e^{\I z e^{\gamma(t-s)}}-1$, we have $f(s,0)=0$ and $\frac{\partial f(s,0)}{\partial z}= {\rm i }e^{\gamma(t-s)}$. If we also make  $g(s)= \nu X(s)$, then from the integral expression for $\Psi$ and  [\ref{eq:3}] we find: 

\beq
\label{eq:mean_generic}
{\rm E}[Y(t)]= -\I \,\frac{\partial \Psi(0;t)}{\partial z} = \int_0^t \nu X(s) e^{\gamma(t-s)} {\rm d}s
\eeq

{ \flushright $\qedsymbol$\\ }

%
%
\bigskip

\noindent {\it {\sc Proposition S3.} ``In the Luria-Delbruck formulation, when the wild type population is modeled with Eq.~{\bf 3}, we have:

\beq
\label{eq:avg_LC_wl_equal}
{\rm E}[Y(t)]_{\rm wl} =  \frac{\nu e^{(2\bar q-1)t} k}{2\bar q} + \nu( t-  k/(2\bar q))X(t;\bar q) -
\frac{\nu e^{-t}(2\bar qt)^k}{2\bar q\,\Gamma(k)}  \qquad  \mbox{\rm for } \gamma = 2\bar q -1
\eeq

\beq
\label{eq:avg_LC_wl_different}
{\rm E}[Y(t)]_{\rm wl} = 
\frac{\nu}{(2\bar q-1)-\gamma}
\left[
X(t;\bar q)
+ \left( \frac{2\bar q}{1+\gamma}\right)^k  \left( e^{\gamma t}- X(t; {\textstyle \frac{\gamma+1}{2}} ) \right)
 -  e^{\gamma t} 
\right]  \qquad  \mbox{\rm for } \gamma \neq 2\bar q -1."
\eeq}\\


\noindent{\it Proof:} Substituting Eq.~{\bf 3} in the article for $X(s)$ in [{\ref{eq:mean_generic}] we have:

\beq
\label{eq:D_L-expected-wl-generic}
	{\rm E}[Y(t)]_{\rm wl} = \frac{\nu e^{\gamma t}}{\Gamma(k)} \int_0^t e^{(2\bar q-1-\gamma)s}\Gamma(k,2\bar qs) {\rm d}s
\eeq

For the special case where the growth rate of both populations are equal ($\gamma= 2\bar q-1$), integrating [\ref{eq:D_L-expected-wl-generic}] and simplifying we find [\ref{eq:avg_LC_wl_equal}]. Expressions of the form $\int e^{ax}\Gamma(k,bx) {\rm d}x$ will show often in our calculations. If $a\neq b$, a closed expression for this antiderivative can be found using integration by parts and the fact that $\frac{{\rm d} }{{\rm d} x}\Gamma(k,x)=-x^{k-1}e^{-x}$ :

\beq
\label{eq:integration-formula}
\int e^{ax}\Gamma(k,bx) {\rm d}x =  e^{ax}\Gamma(k,bx)/a -\left(\frac{b}{b-a} \right )^k \Gamma(k,(b-a)x)/a
\eeq

\noindent We can use [{\ref{eq:integration-formula}] to find ${\rm E}[Y(t)]_{\rm wl}$ when $\gamma \neq 2\bar q-1$. In this case ${\rm E}(Y(t))_{\rm wl}$ is given by Eq.~{\ref{eq:avg_LC_wl_different}. 

{ \flushright $\qedsymbol$\\ }

%
%
\bigskip 

\noindent{\it {\sc Proposition 4.} ``In the Luria-Delbruck formulation, when the wild type population is modeled with Eq.~{\bf 3}, we have:

\beq
\label{eq:LD-var-wl}
{\rm V}[Y(t)]_{\rm wl}=  \left\{ \begin{array}{ll}
\frac{\nu}{2\bar q-1-2\gamma}
\left[
X(t;\bar q)
+ \left( \frac{2\bar q}{1+2\gamma}\right)^k  \left( e^{2\gamma t}- X(t; {\textstyle \frac{2\gamma+1}{2}} ) \right)
 -  e^{2\gamma t}
\right] & \quad (2\gamma \neq 2\bar q-1)  \\
\frac{\nu e^{(2\bar q-1)t} k}{2\bar q} + \nu( t-  k/(2\bar q))X(t;\bar q) -
\frac{\nu e^{-t}(2\bar qt)^k}{2\bar q\,\Gamma(k)} & \quad (2\gamma = 2\bar q-1)"
       \end{array} \right. 
\eeq}\\


\noindent{\it Proof:} From [{\bf 5}] the characteristic function of the number of mutants $\Psi(z;t)= \exp \left\{ \int_0^t \nu X(s) [\psi(z;t,s) -1] {\rm d}s \right\}$, where $\psi(z;t,s)= e^{\I z e^{\gamma(t-s)}}$. Let $f(s,z)$ and $g(s)$ be two functions with continuous second order partial derivates, then:

\beq
\label{eq:second_derivative}
\frac{\rm d^2}{{\rm d}z^2} \exp\left[\int_0^t g(s)f(z,y) {\rm d}s \right] = e^{\int_0^t g(s)f(s,z){\rm d}s} 
\left\{ \int_0^t g(s)\frac{\partial^2 f(s,z)}{\partial z^2}{\rm d}s + \left( \int_0^t g(s) \frac{\partial f(s,z)}{\partial z} {\rm d}s  \right)^2 \right\}
\eeq

If we make again $g(s)= \nu X(s)$ and $f(s,z)= e^{\I z e^{\gamma(t-s)}}-1$, then $\frac{\partial^2 f(s,0) }{\partial z^2}= - e^{2\gamma(t-s)}$, and substituting this expression into (\ref{eq:second_derivative}) we find:

\beq
\label{eq:var_LC_generic}
{\rm V}[Y(t)]= (-{\rm i})^2 \frac{\partial^2 \Psi(0;t)}{\partial z^2} - {\rm E}[Y(t)]^2 = \int_0^t \nu X(s) e^{2\gamma(t-s)}{\rm d}s
\eeq

When wild type cells have replicative limits, the equations for the variance in [\ref{eq:LD-var-wl}] can then be deduced by comparing [\ref{eq:mean_generic}] and [\ref{eq:var_LC_generic}] and using the expressions for the expected value in Proposition 3.

{ \flushright $\qedsymbol$\\ }

%
%

\subsection*{Lea-Coulson formulation}

%
%

\noindent{\it {\sc Proposition 5.} ``With and without limits: {\rm I.} The expected value in the Luria-Delbr\"uck formulation and the Lea-Colusion formulation are equal, i.e.: ${\rm E}[Y(t)]^{\scriptstyle (LD)} = {\rm E}[Y(t)]^{\scriptstyle (LC)}$. {\rm II.} The variances in the Luria-Delbr\"uck formulation and the Lea-Colusion formulation satisfy the identity:}

\beq
\label{eq:var-relation} \textstyle
{\rm V}[Y(t)]^{\scriptscriptstyle \rm (LC)}=
{\rm E}[Y(t)] +
\frac{2\alpha}{\alpha-\beta} \left( {\rm V}[Y(t)]^{\scriptscriptstyle \rm (LD)} - {\rm E}[Y(t)] \right)
\eeq\\


\noindent{{\it Proof:}} I. From the introduction in the article of the Lea-Coulson formulation, the probability generation function of the number of mutants $Y(t)$ equals:

\beq
\label{eq:pgf_LC_general}
\Phi(z;t)= \exp \left\{ \int_0^t \nu X(s) [\phi(z;t,s) -1] {\rm d}s \right\}
\eeq

\noindent where  $\phi(z;t,s)$ is given by [19]. Given that ${\rm E}[Y(t)]= \frac{\partial}{\partial z} \Phi(1;t)$, we can use [\ref{eq:pgf_LC_general}] to calculate  the expected number of mutants as a function of time.  Let us use Eq.~\ref{eq:3} and substitute in  this expression $g(s)=\nu X(s)$ and $f(s,z)= \phi(z;t,s)-1$. We have $f(s,1)=0$ and $\frac{\partial f(s,1)}{\partial z} = e^{(\alpha-\beta)(t-s)}$.  The expected number of mutants is then equal to:

\beq
	{\rm E}[Y(t)]= \int_0^t \nu X(s)e^{(\alpha-\beta)(t-s)} \, {\rm d}s
\eeq

By writing $\gamma= \alpha-\beta$ (the net growth rate of the mutant population) in equation [\ref{eq:mean_generic}] we find that the respective expected values (with or without replicative limits) are the same for the LD and LC formulations.\\


II. We can also calculate the variance from the p.g.f. by means of the identity ${\rm V}[Y(t)] = \frac{\partial^2}{\partial z^2} \Phi(1;t) + {\rm E}[Y(t)] - {\rm E}[Y(t)] ^2$. Again making $g(s)=\nu X(s)$ and $f(s,z)= \phi(z;t,s)-1$ and using equations [\ref{eq:3}] and [\ref{eq:second_derivative}], we find:

\beq
 {\rm V}[Y(t)] = \int_0^t \nu X(s) \frac{\partial^2 \phi(1;t,s)}{\partial z^2} {\rm d}s + {\rm E}[Y(t)]
\eeq

\beq
\label{eq:smth}
	= - \frac{2\alpha}{\alpha-\beta} \int_0^t  \nu X(s) e^{2(\alpha-\beta)(t-s)} \left[ e^{-(\alpha-\beta)(t-s)} -1 \right] {\rm d}s + {\rm E}[Y(t)]
\eeq

\noindent Substituting $\gamma$ by $\alpha-\beta$ in Eqs.~\ref{eq:mean_generic} and \ref{eq:var_LC_generic}. It follows that the integrand in [\ref{eq:smth}] equals ${\rm E}[Y(t)] - {\rm V}[Y(t)]^{\scriptstyle \rm (LD)}$ proving identity [\ref{eq:var-relation}].

{ \flushright $\qedsymbol$\\ }

%
%

\subsection*{Distribution}

%
%

\noindent{\it {\sc Proposition 6.} ``If $\Phi$ is given by equation {\rm [\ref{eq:pgf_LC_general}]},  $\sum a_n(t,s)z^n$ is the power series representation of $\phi(z;t,s) -1$ centered around $z=0$, and $X(s)$ is given by {\rm[{\bf 3}]}, then $\int_0^t \nu X(s) \sum a_n(t,s)z^n {\rm d}s = \sum \int_0^t \nu X(s) a_n(t,s)z^n {\rm d}s$."}\\

{\it Proof:} The interchange is licit when the growth of the wild type population is exponential, i.e. when $X(t)=e^{\delta t}$ for any $\delta>0$ (reference \cite{Stewart:1990kx}). Hence, since $X(t;\bar q) \leq e^{(2\bar q-1)t}$, it follows directly from the dominated convergence theorem that the interchange is also licit when the number of wild type cells is modeled as $X(t;\bar q)$.

{ \flushright $\qedsymbol$\\ }

%
%

\subsection*{Bartlett's formulation}

%
%

\noindent{\it {\sc Proposition 7.}  ``If $X(t)$ is modeled by Eq.~3, then the  joint p.g.f. $\Phi$, of the number of wild type cells  and mutants  satisfies the differential equation:

 \beq
 \label{eq:diff_eq_bartlett_wl}
 \frac{\partial \Phi}{\partial t}=  
 \sum_{j=1}^k \left[  q (1-\mu) z_{j-1}^2 + q \mu z_{j-1} w - z_j  + (1-q) \right] \frac{\partial \Phi}{\partial z_j}
 +
(1-q)(1-z_0) \frac{\partial \Phi}{\partial z_0}
 +
 \left( \alpha w^2 - (\alpha+\beta) w + \beta \right) \frac{\partial \Phi}{\partial w}
 \eeq
 }
 
 
 \noindent{\it Proof:} Bartlett's formulation defines a continuous-time Markov process. Let $x_j(t)$ be the number of cells with replication capacity $j$ at time $t$, and $y(t)$ the number of mutants. Also, let $P(t)= {\rm Prob}(x_0(t)=i_0,\dots,x_k(t)=i_k,Y=l)$ be the probability that for all $0\leq j \leq k$, $x_j(t) = i_j$ and $y(t)=l$. As a matter of notation we can then write any other probability by considering only the indices that change with respect to $P(t)$. We can then write the Kolmorov forward equations by consider all the possible changes in the system. The possible events are described by the quantities $L1$-$L7$ below.\\

 $L1$: (plus) \mbox{$x_j$ cell divides producing two $x_{j-1}$ cells to enter state}

 \[
 L_1 =  (1-\mu) q  \sum_{j=1}^k  (i_j+1){\rm Prob}(x_{j-1}(t) = i_{j-1}-2, x_{j}(t) = i_j+1)
  \]

 $L2$: (plus) \mbox{$x_j$ cell divides producing one $x_{j-1}$ cell and one $y$ cell to enter state}.

 \[
 L_2 =  \mu q  \sum_{j=1}^k (i_j+1){\rm Prob}(x_{j-1}(t) = i_{j-1}-1, x_{j}(t) = i_j+1, y(t)=l-1)
 \]
 
  $L3$: (minus) \mbox{$x_j$ cell divides to exit state}.	$L_3 =   q  \sum_{j=1}^k (i_j) {\rm Prob}(x_{j}(t) = i_j) $.\\

  $L4$: (minus) \mbox{$x_j$ cell dies to exit state}.		$ L_4 =   (1-q)  \sum_{j=0}^k (i_j) {\rm Prob}(x_{j}(t) = i_j)$.\\
 
  $L5$: (plus) \mbox{$y$ cell divides to enter state}.		$L_5 =   \alpha  (l-1){\rm Prob}(y(t)= l-1)$.\\
 
 $L6$: (plus) \mbox{$y$ cell dies to enter state}.		$L_6 =   \beta (l+1){\rm Prob}(y(t)= l+1)$\\
 
  $L7$: (minus) \mbox{$y$ divides or dies to exit state}.	$ L_7 =   (\alpha+\beta){\rm Prob}(y(t)= l)$.\\
  
  The Kolmogorov forward equation is then: $\dot P = L1 + L2 - L3 -L4 + L5 +L6 - L7$. Multiplying both sides by $(\prod_{j=0}^k z_j^{i_j})w^l$ and summing over the vector $(i_0,\dots,i_k,l)$ we find the partial differential equation [\ref{eq:diff_eq_bartlett_wl}] for the joint p.g.f. $\Phi(z_0,\dots,z_k,l; t)$.
  
   { \flushright $\qedsymbol$\\ }

%
%

	\bibliographystyle{pnas}
	\bibliography{references-LD}